\documentclass[10pt,twoside]{hsqcd}
\usepackage{epsf,amsmath}
\usepackage{graphicx}
\usepackage{cite}

\begin{document}

\title{PROTON  STRUCTURE
\thanks{presented at
HSQCD2005,  20 - 24th September, St. Petersburg.}
}

\author{\underline{J\"org Gayler} \\ \\
{\it for the H1 and ZEUS Collaborations} \\ \\
DESY \\
 }

\maketitle

\begin{abstract}
\noindent Recent inclusive neutral current and charged current DIS data from
 HERA are discussed in context of
 pQCD and parton density functions.
\end{abstract}




\section {Introduction}

Since the early days of Quantum-Chromo-Dynamics (QCD), deep inelastic  
scattering (DIS) played a decisive role in stimulating theoretical ideas
and testing them experimentally. Now the theory is well
established and perturbative QCD (pQCD) is  
well tested in many processes, but the details of lepton
 nucleon interactions, in particular at small Bjorken $x$, and the structure of the proton 
in terms of parton densities (pdfs) are still objects of intensive research,
both, experimentally and theoretically. The understanding of proton structure is of fundamental interest, but also of great practical importance
for the quantitative analysis of present and future hadron hadron scattering 
experiments like those at the forthcoming LHC.

Since its beginning in 1992, the HERA accelerator has been a major source 
of information on  proton structure. The fully inclusive
neutral current (NC) and charged current~(CC) reactions
($ep \rightarrow eX$ and $ep \rightarrow \nu X$, respectively) are particularly
suited to unveil the parton densities of the proton
due to the comparatively simple and well understood theoretical description
of these reactions.

However, to obtain pdfs at large $x$ and for the decomposition of $u$ and $d$ quark flavours, HERA data are usually combined with results of  other 
experiments, which leads to additional systematic uncertainties. In particular
it is desirable to determine the $d$ quark density by the HERA experiments
alone, free of nuclear corrections.

This paper deals with collinear pdfs, which contain no information
on parton transverse momenta and parton-parton  correlations.
Constraints on general parton densities (GPDs) are deduced
from some exclusive final states (see~\cite{proc}), while 
 the constraints on  collinear pdfs are obtained from
inclusive scattering.
After some description of the kinematics and
relations of
cross sections and structure functions (section~\ref{sec:formal}),
I first present 
shortly the most recent
data in CC interactions collected by HERA II, the upgraded HERA machine 
 which provides 
 longitudinally polarised electron and
positron beams at the H1 and ZEUS experiments (section~\ref{sec:hera2}).  
Sections~\ref{sec:nc} and ~\ref{sec:cc} present recent NC and CC data which 
partially have been used already, for pdf fits, which are described in 
section~\ref{sec:fits}. The measured contributions of charm and
 beauty production
to the proton structure function is shortly discussed
 in section~\ref{sec:beau},
followed by conclusions.

\section{Cross Sections and Structure Functions}
\label{sec:formal}

In  inclusive $ep$ scattering the proton structure can be probed
 by $\gamma$ or $Z^0$ exchange, i.e.
 by neutral current  interactions,
 or by $W$  exchange, i.e.
 by charged current  interactions.
 The NC differential cross section can be expressed in terms of
 three structure functions, $\tilde{F_2}$, $\tilde{F_3}$ and $\tilde{F_L}$:  
\begin{equation}
d^2 \sigma^{\pm}_{NC}/dxdQ^2 =
\frac {2\pi \alpha^2}{xQ^4} [ Y_+ \cdot \tilde{F_2}
         \;   \mp \; Y_- \cdot x\tilde{F_3}  -y^2
      \cdot \tilde{F_L}] \equiv \frac {2\pi \alpha^2}{xQ^4} Y_+
       \tilde{\sigma}^{\pm}_{NC}\;,
\end{equation}
where  $Y_{\pm} = 1 \pm (1 - y)^2$. Here, $Q^2=-q^2$, with $q$ being
 the four-momentum
 of the exchanged gauge boson,  $x = Q^2/2(P\cdot q)$,
 the momentum fraction of the proton carried by the parton
 participating in the interaction, and
 $y = (P\cdot q)/(P\cdot k)$,
 the
 inelasticity, where $k (P)$ is the
 four-momentum of the incident electron (proton).
 The term $\tilde{\sigma}^{\pm}_{NC}$ is called ``reduced cross section".
 It is a combination of structure functions directly  related to the
 differential cross section.
 The structure function $\tilde{F_2}$ is the dominant contribution in
 most of the phase space and can, in leading order (LO) QCD,
  be written in terms of
 the quark densities
 $\sim x \sum_q e_q^2(q(x) + \bar q(x))$.
 The term  $x\tilde{F_3}$, which would vanish in the absence of $Z^0$ 
 exchange, contributes 
  significantly
  only at at high $Q^2$,
 and is to LO  $\sim  x \sum_q (q(x) -\bar q(x))$, that is,
  it is given by the
 valence quarks.
The longitudinal
 contribution $\tilde{F_L}$ is important in Eq. (1)  at large $y$.
 At small $x$, to order $\alpha_s$,  $\tilde{F_L} \sim \alpha_s g$,
 where $g$ is the gluon density.
 
 Similarly, the CC cross section can be written
\begin{equation}
d^2 \sigma^{\pm}_{CC}/dxdQ^2 =
\frac {G^2_F}{2\pi x} (\frac{M^2_W}{Q^2 + M_W^2})^2
  \cdot    \tilde{\sigma}^{\pm}_{CC}\;\;,
\end{equation}
where $G_F$ is the Fermi coupling constant.
 
In LO
$\;\;\;\tilde{\sigma}^+_{CC} = x[(\bar u(x) + \bar c(x)) + (1 - y)^2(d(x) + s(x))]$\\
 \hspace*{32pt} and
$\;\;\;\tilde{\sigma}^-_{CC} = x[(u(x) + c(x)) + (1 - y)^2(\bar d(x) + \bar s(x))]$.\\
 The $d$-quark density is therefore directly accessible
 in $e^+p \rightarrow \bar \nu_e X$ scattering,
 avoiding the nuclear corrections necessary when electron 
 deuteron scattering is used.

\section{Most Recent HERA  Results in inclusive DIS}
\label{sec:hera2}

\begin{figure}[htb] \unitlength 1pt
   \begin{center}

   \begin{picture}(170.,140.)
    \put(-100,-18){\includegraphics[width=0.4\textwidth]{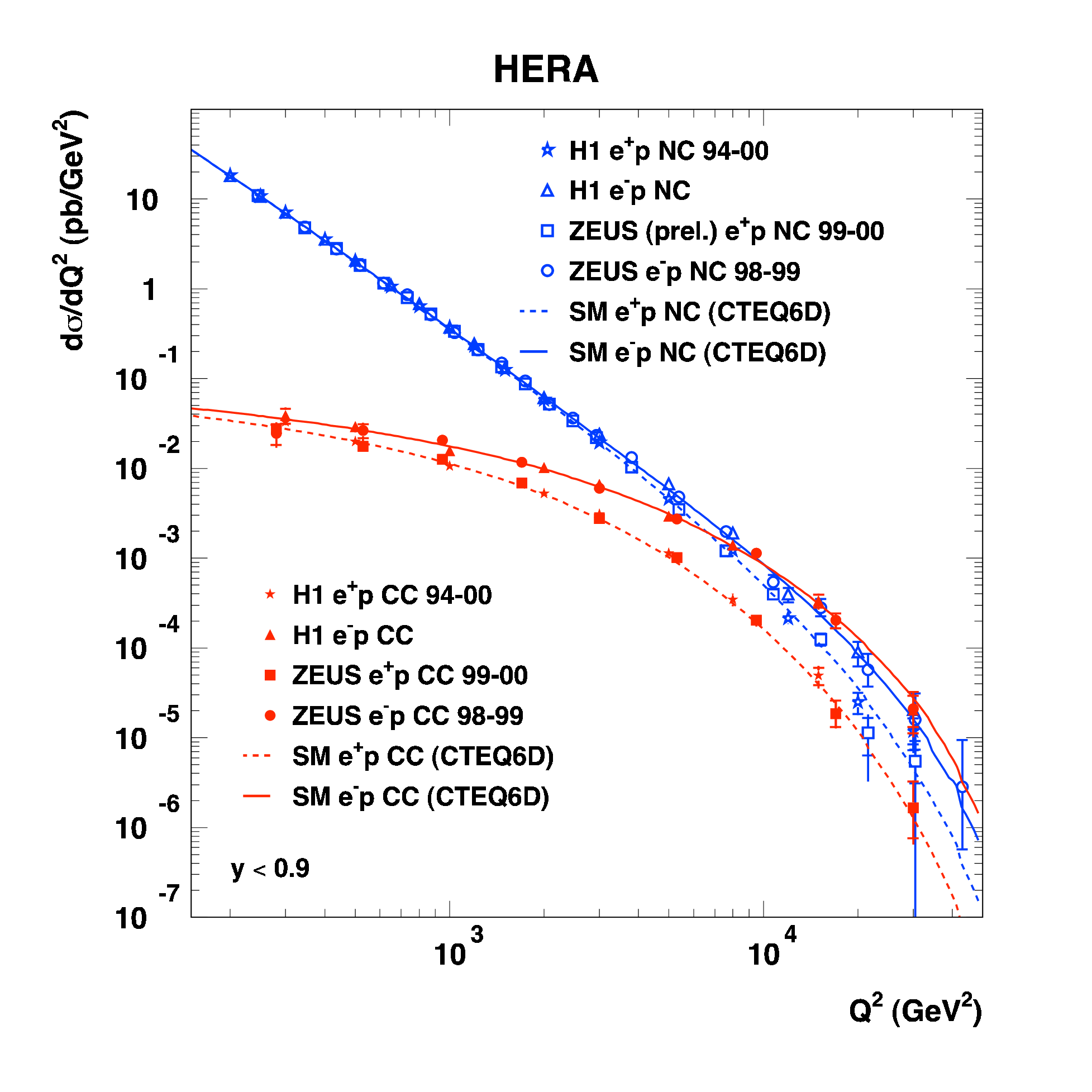}}
    \put(100,-6){\includegraphics[width=0.408\textwidth]{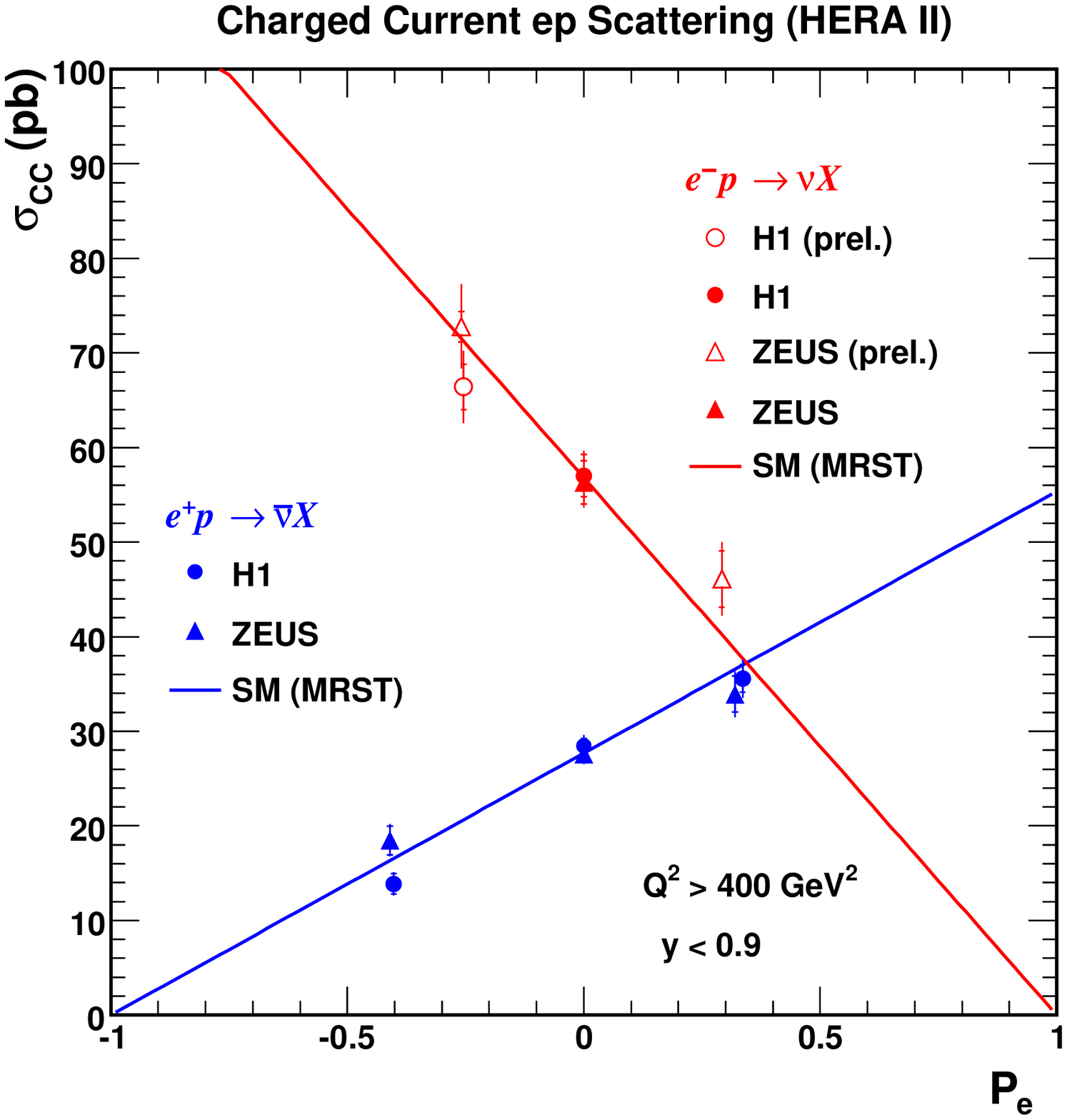}}
   \end{picture}
   \caption{NC and CC cross sections vs. $Q^2$ (left).
   Polarisation dependence of the total CC cross section (right).
}
   \label{fig:ccpol}
   \end{center}
   \end{figure}

HERA covers a large phase space in $x$ and $Q^2$ well suited for pdf analyses,
 but also for the analysis of electro-weak effects in DIS. Fig. \ref{fig:ccpol} (left) shows 
 NC and CC cross sections spanning three orders of magnitude in  $Q^2$,
  which get similar magnitude at $Q^2 \sim 10^4$ where the propagator mass
  is in the range of $M_W   ^2$. The data thus illustrate electro-weak 
  unification. The standard model (SM) is further tested by the
  dependence of the CC cross section on the polarisation of the primary
  electron or positron.
  The extrapolated cross sections for 
  for fully left handed positrons
  are  consistent with zero, the SM expectation
  (Fig. \ref{fig:ccpol}, right).
  Thus, there is no indication of right handed weak currents (see \cite{ri}
  for further details). For a recent combined electro-weak and pQCD
  analysis, see~\cite{Aktas:2005iv}.
  
\section{NC data}
\label{sec:nc}

\begin{figure}[htb] \unitlength 1pt
   \begin{center}
   \begin{picture}(150.,172.)
    \put(-100,-6){\includegraphics[width=0.36\textwidth]{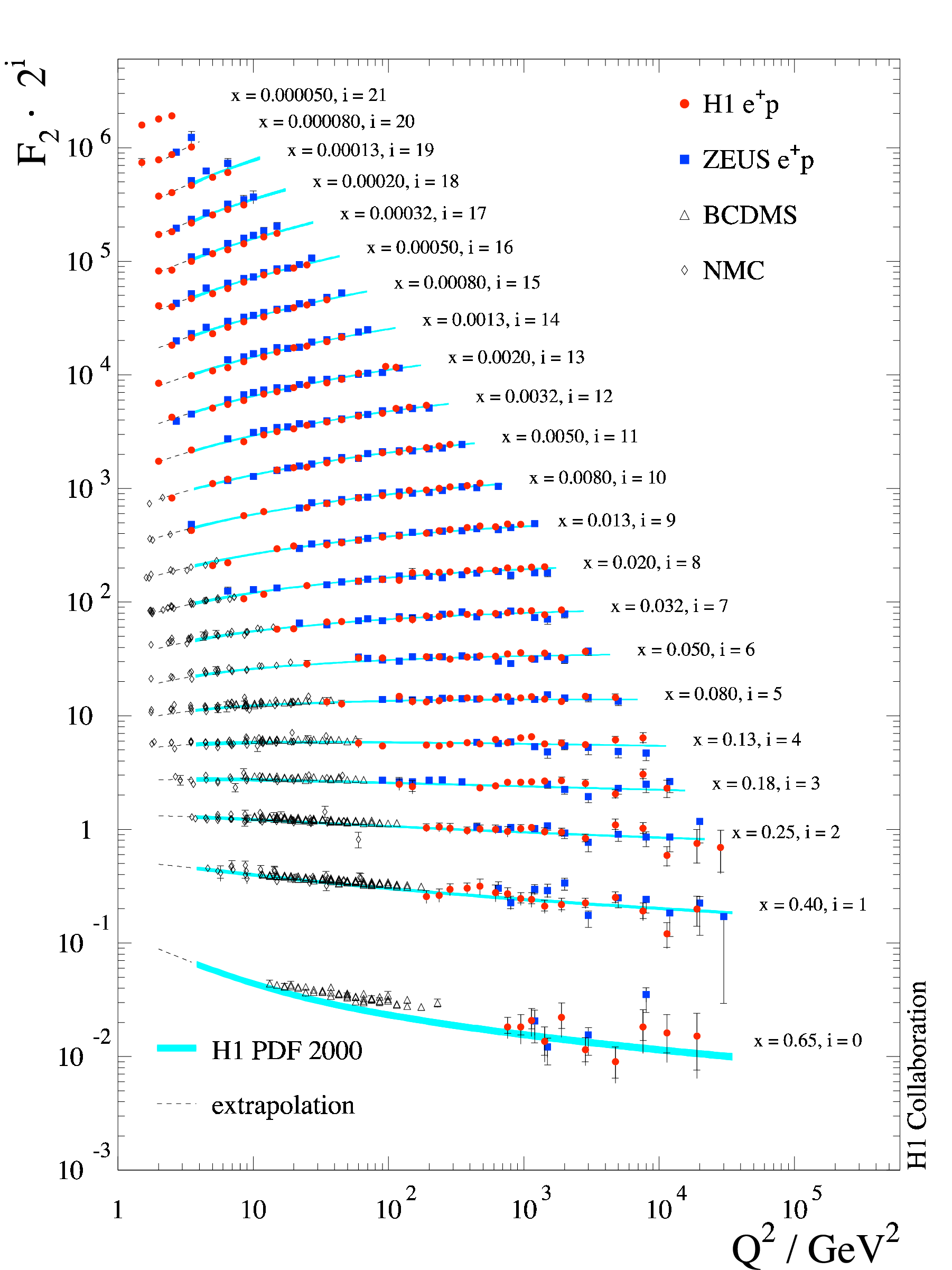}}
     \put(55,-50){\includegraphics[width=0.45\textwidth]{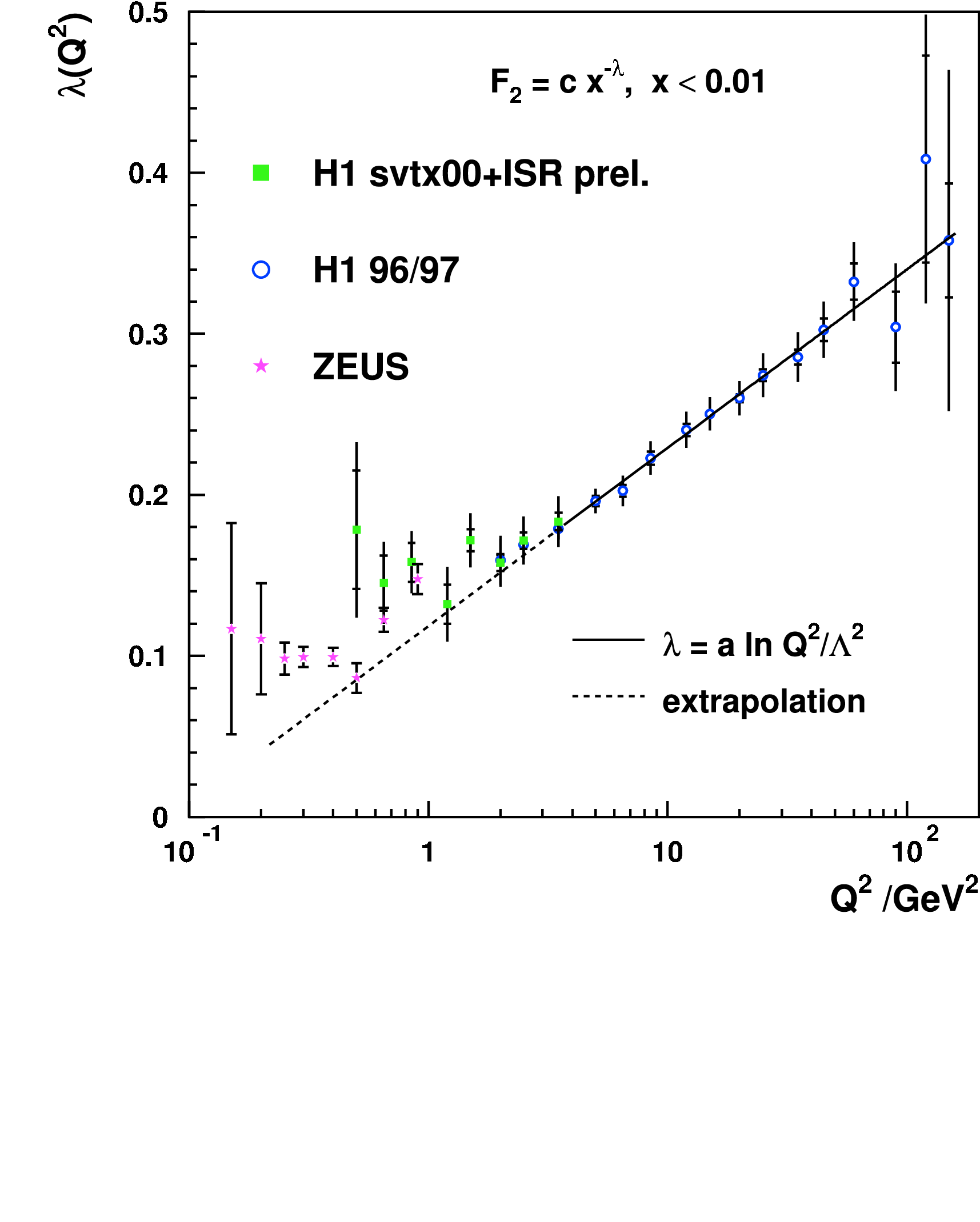}}
   \end{picture}
   \caption{$F_2$ vs. $Q^2$ (left).
    $\lambda (Q^2) $ from fits $F_2 = c(Q^2) \cdot x^{-\lambda (Q^2)}$ 
    for $x < 0.01$ (right).
}
   \label{fig:ncvsq2}
   \end{center}
   \end{figure}

 The HERA I inclusive NC $F_2$ data are shown in Fig.~\ref{fig:ncvsq2}. 
  They span 4 orders of magnitude in $x$ and $Q^2$
 and are closing the gap to the fixed target data.
  At small  $x$ ($x < 0.01$),
  the rise of the structure function towards small $x$ 
  can well be described by  $F_2 = c(Q^2) \cdot x^{-\lambda}$
  with $\lambda $ logarithmically rising with $Q^2$
  (Fig.~\ref{fig:ncvsq2}, right)
   and $c(Q^2)$ being roughly constant
  for  $Q^2 > 3$ GeV$^2$.
    At small $Q^2$, $\lambda$ deviates from this
   logarithmic dependence, consistent with   $\lambda \rightarrow 0.08$ for
    $Q^2 \rightarrow 0$, as expected from soft hadronic interactions.

\begin{figure}[htb] \unitlength 1pt
   \begin{center} 
   \begin{picture}(170.,116.)
    \put(-100,-10){\includegraphics[width=0.30\textwidth]{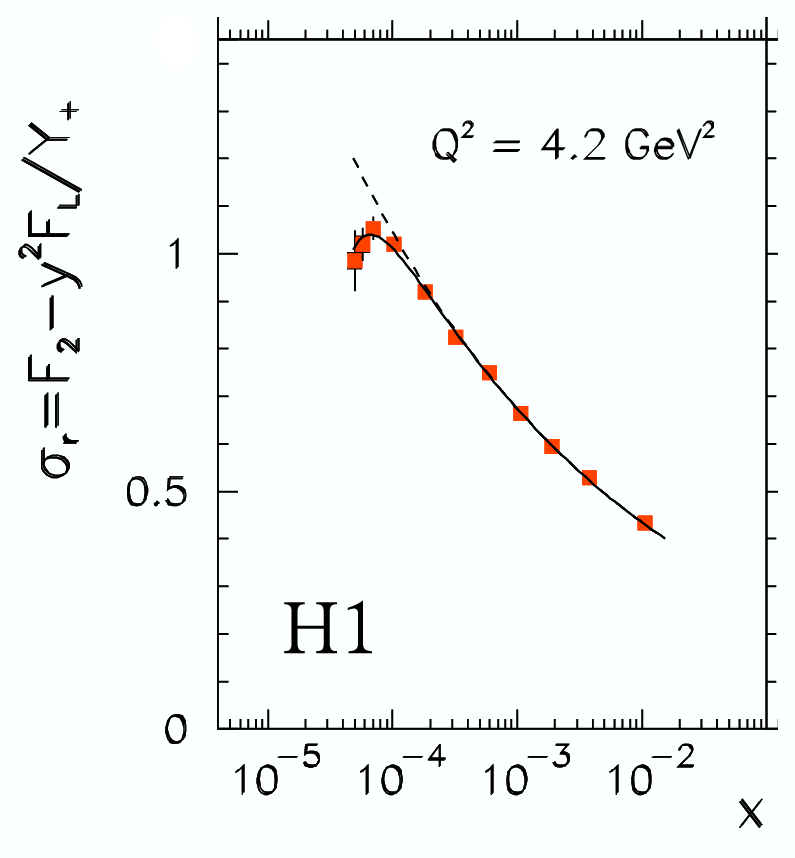}}
    \put(45,-4){\includegraphics[width=0.50\textwidth]{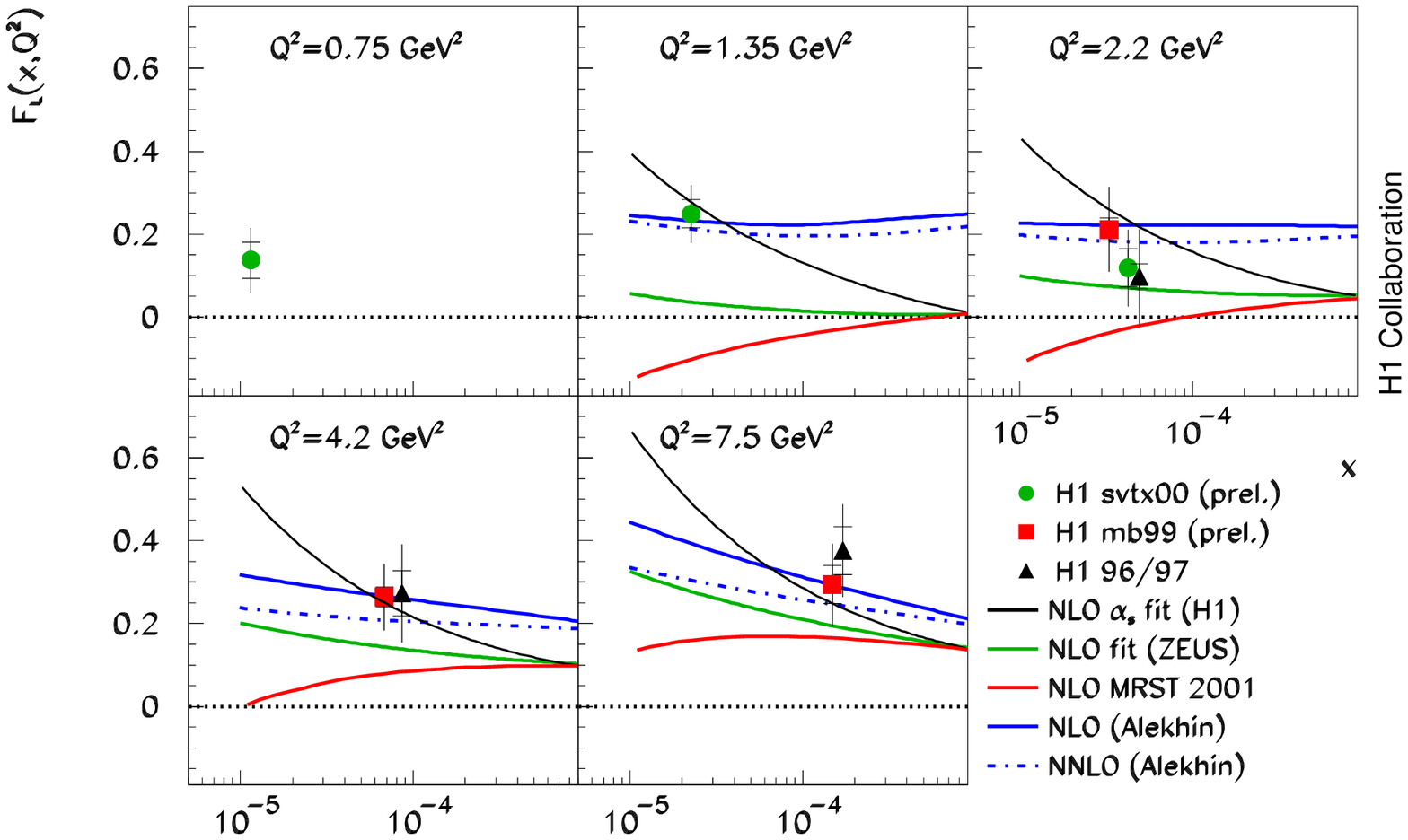}}
   \end{picture}
   \caption{Left: Reduced cross section vs. $x$ for $Q^2 = 4.2$ GeV$^2$.
    Solid line: fit $\sigma _r = c \cdot x^{-\lambda } - y^2/Y_+F_L$, broken
   line: $c \cdot x^{-\lambda }$. Right: $F_L(x,Q^2)$ 
   for fixed $Q^2$~\cite{H1prelim-03-043} and theoretical
    predictions based
     on ~\cite{Adloff:2000qk,Chekanov:2002pv,Martin:2001es,Alekhin:2002fv}   
}
   \label{fig:sigmal}
   \end{center}
   \end{figure}

   The longitudinal proton structure function $F_L$ is of considerable
   interest due to the relation with the proton gluon density,
   $F_L \sim \alpha_s(Q^2)g(x,Q^2)$.
   However no direct measurements of  $F_L$
   are available in the  HERA regime. Such measurements require changes of 
   the center of mass energy 
 $\sqrt s$, i.e. of the beam energies. 
 However at very high $y$, the signature of $F_L$ is a drop of the reduced 
 cross section, because the longitudinal contribution vanishes
 at $y = 1$. This is visible in Fig.~\ref{fig:sigmal}, left,
 for an $x$~distribution at
 $Q^2 = 4.2$ GeV$^2$~\cite{H1prelim-03-043}
 \footnote{It is therefore important to include in pQCD analyses
 the available reduced cross sections at high $y$
 and not $F_2$ results only.}. 
 The H1 collaboration used various methods with assumptions in the 
 framework of pQCD to exploit this effect.
The results  are presented in Fig.~\ref{fig:sigmal}, right, where 
the determined $F_L$ values are compared with predictions based
on the   
NLO fits of H1~\cite{Adloff:2000qk}, ZEUS~\cite{Chekanov:2002pv} and
MRST21~\cite{Martin:2001es} and the 
NLO and NNLO analyses of S.~Alekhin~\cite{Alekhin:2002fv}. 
The spread of the predictions is large  and shows
the importance of direct $F_L$ measurements. 
  
\begin{figure}[htb] \unitlength 1pt
   \begin{center}
   \begin{picture}(150.,145.5)
    \put(-60,-8){\includegraphics[width=0.63\textwidth]{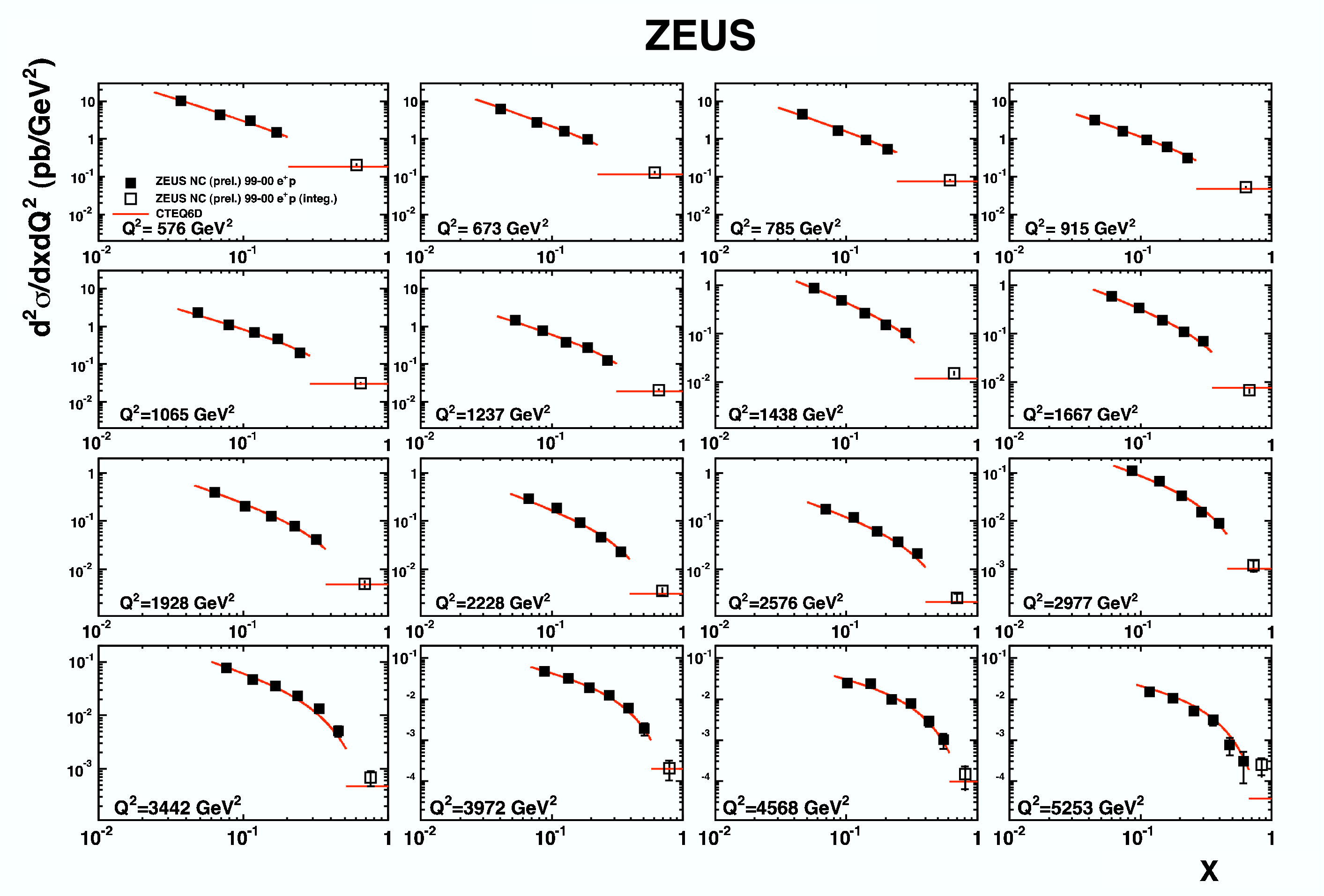}}
   \end{picture}
   \caption{$d\sigma /dx dQ^2$ for NC $e^+p$ compared to SM 
    expectation (CTEQ6D pdfs, solid line).
}
   \label{fig:zeushighx}
   \end{center}
   \end{figure}

  At high $x$ the uncertainties are still quite large
  (see Fig.~\ref{fig:ncvsq2}, left, $x = 0.65$).
  This led the ZEUS collaboration~\cite{zeushighx}
 to apply a new technique, which exploits
  the fact, that at very high $x$, where $Q^2$ can still be well measured 
  using the scattered electron,
  the hadronic jet
   vanishes in the beam pipe at small forward angles 
   due to kinematics. If this happens,
  $x$ is above some $x_{edge}$ depending on  $Q^2$.
  Fig.~\ref{fig:zeushighx} shows $x$ distributions in bins of $Q^2$ from
  $Q^2 = 576$ GeV$^2$ to  $Q^2 = 5253$ GeV$^2$. In each case, the highest   
  $x$~bin 
  is an average over the  $x$ range where the jet vanishes
   in the beam pipe. 
  These data can provide significant constraints at high $x$ in pQCD analyses.

\section{CC data}
\label{sec:cc}

\begin{figure}[htb] \unitlength 1pt
   \begin{center}  
   \begin{picture}(150.,160.)   
    \put(-40,-20){\includegraphics[width=0.44\textwidth]{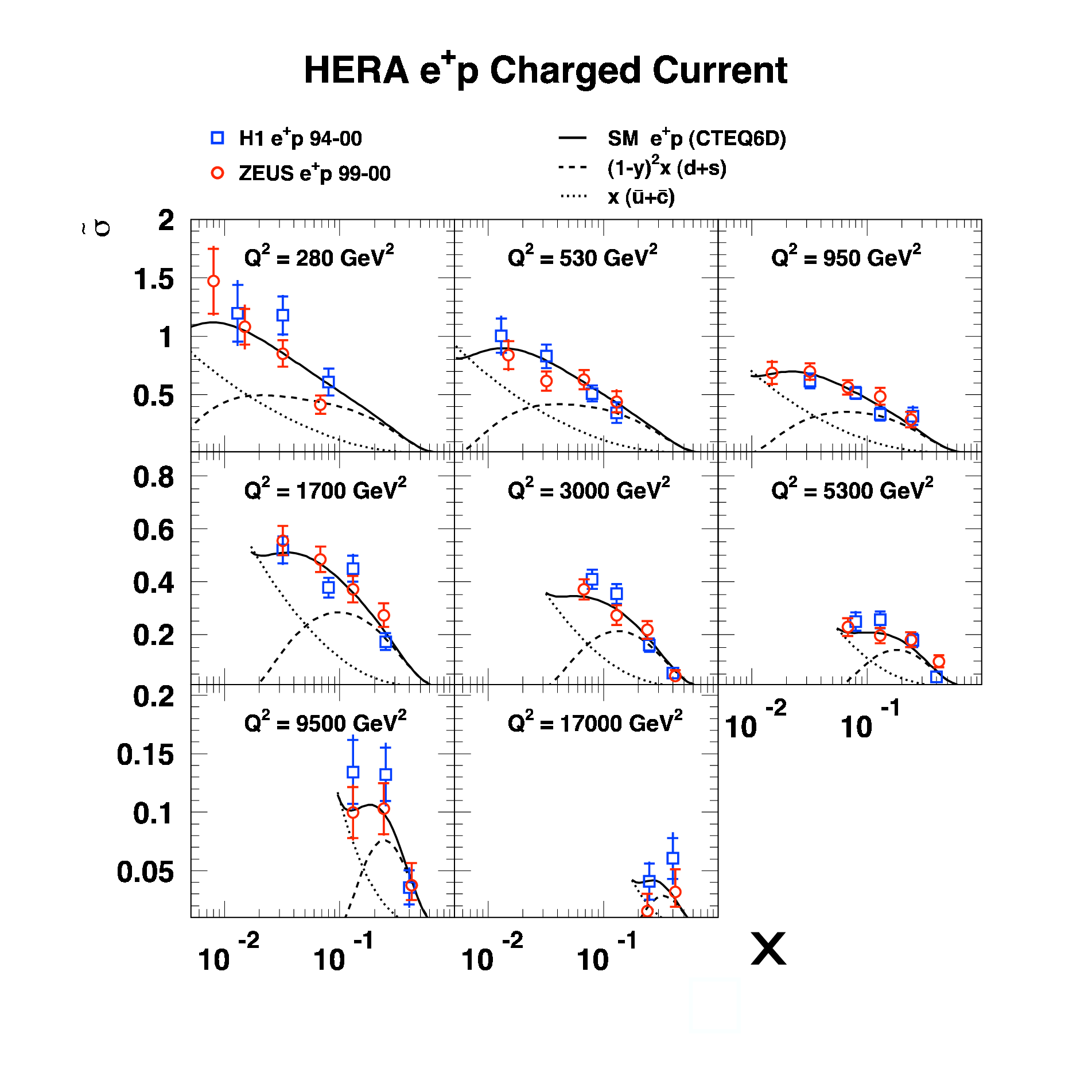}}
   \end{picture}
   \caption{Reduced CC cross sections of H1 and ZEUS as function of 
   $x$ and $Q^2$ compared with predictions based on CTEQ6D (solid line),
   showing separately the contributions of  $d + s$ (dashed) and
    $\bar{u} + \bar{c}$ (dotted).
}
   \label{fig:ccvsx}
   \end{center}
   \end{figure}

  HERA has the unique possibility to determine the $d$-quark density in the
  proton free of nuclear effects by the CC reaction 
   $e^+p \rightarrow \bar{\nu } X$.
  Fig.~\ref{fig:ccvsx} shows $x$ distributions at different $Q^2$ from 
 $Q^2 = 280$ to  $Q^2 = 17000$ GeV$^2$.  
   They are dominated by the $d$-quark for $x > 0.2$ and 
   will, with final HERA statistics, significantly improve
   the knowledge on  the $d$-quark density. Another constraint is provided
   by $e^+$ and $e^-$ NC data at high $Q^2$, because $u$ and $d$ quarks enter
   in 
    $xF_3 \sim 2 u_v + d_v$ with weights which differ from those in $F_2$. 

\section{Pdf Fits}
\label{sec:fits}

 Essentially the available
 information on pdfs at small $x$ is provided by 
 inclusive DIS measurements at HERA,
 showing in particular a strong rise of the sea and gluon pdfs towards small
   $x$.
  In the standard DIS QCD analyses, a parameterisation
 of the pdfs at a starting scale $Q_0^2$ is assumed,
 which are evolved to higher $Q^2$ using the
 NLO DGLAP equations~\cite{Furmanski:1980cm}.
 The parameters at $Q_0^2$ are then determined
 by a fit of the calculated cross sections to the
 data.

\begin{figure}[htb] \unitlength 1pt
   \begin{center}
   \begin{picture}(150.,25.)
    \put(-55,-6){\includegraphics[width=0.09\textwidth]{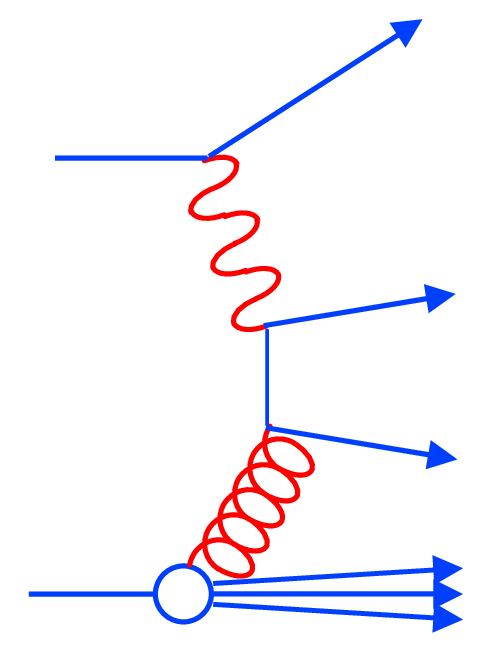}}
    \put(50,-8){\includegraphics[width=0.09\textwidth]{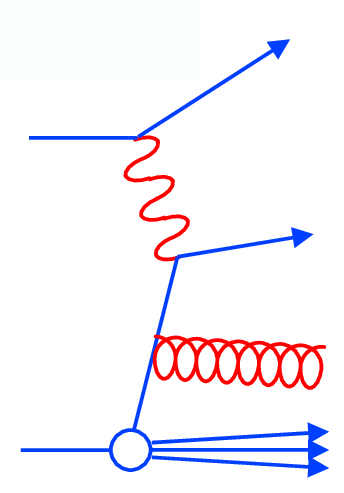}}
    \put(145,-8){\includegraphics[width=0.14\textwidth]{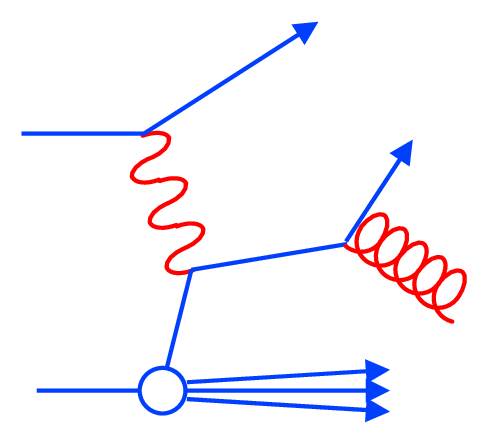}}
   \end{picture}
   \caption{Diagrams for di-jet production. Boson gluon fusion (left),
    QCD-Compton graphs (central and right).
}
   \label{fig:graphs}
   \end{center}
   \end{figure}

The gluon density of the proton at small $x$ is
 quite well determined, although it
 influences the cross sections only at next to leading order (NLO) 
via scaling violations. Di-jet measurements, however, are sensitive to the gluon
density already in leading order by boson-gluon fusion (Fig.~\ref{fig:graphs}, left).
 Further, the 
 production of jets at large
$p_t$, with respect to the exchanged boson direction, implies that large momentum fractions of the proton are involved.
  The correlation of the gluon density with $\alpha _s$ in 
boson-gluon-fusion  is broken by the contributing QCD-Compton
graphs (Fig.~\ref{fig:graphs}, central and right). 

\begin{figure}[htb] \unitlength 1pt
   \begin{center}
   \begin{picture}(150.,172.)
         \put(100,-7){\includegraphics[width=0.33\textwidth]{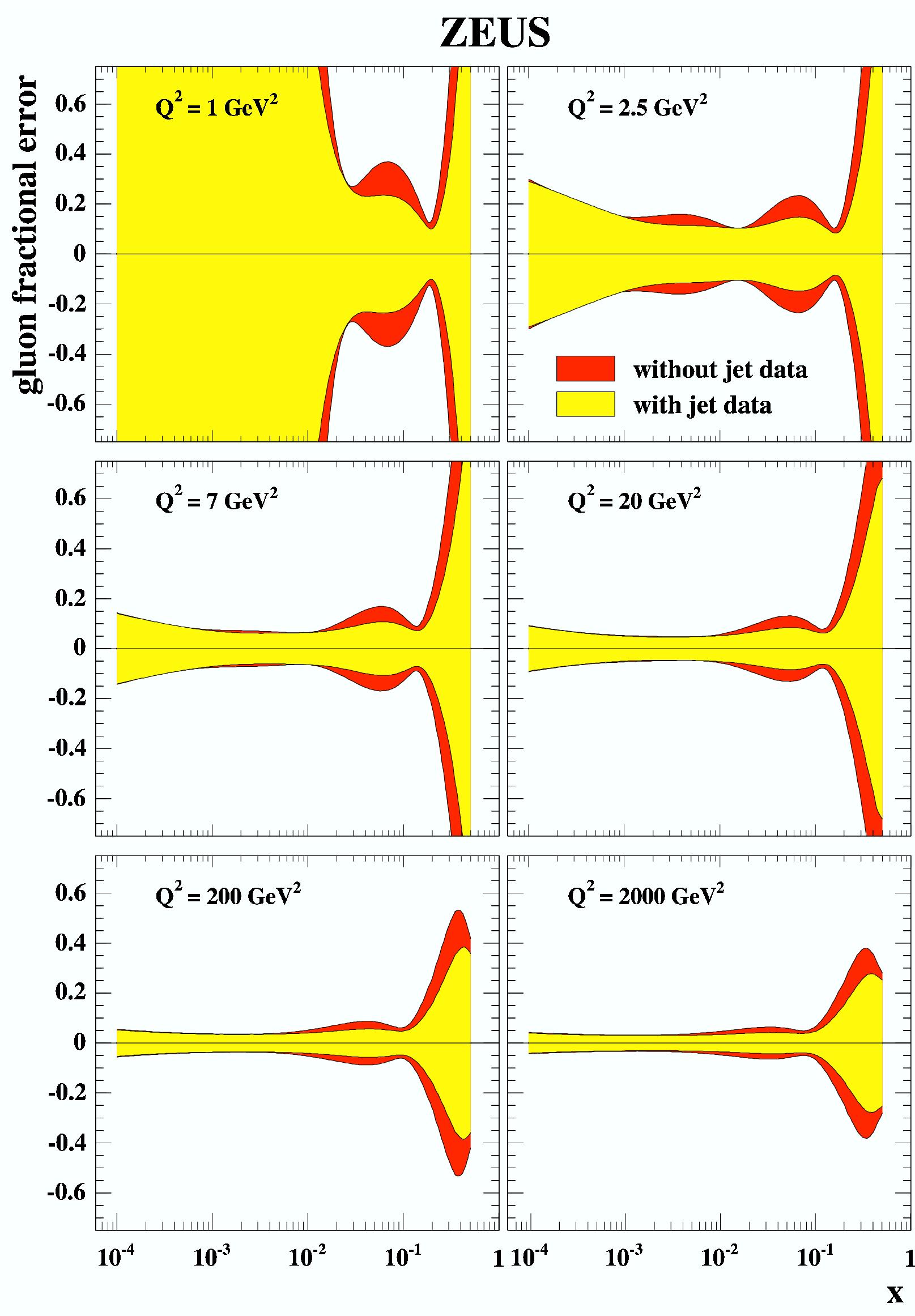}}
     \put(-100,-8){\includegraphics[width=0.365\textwidth]{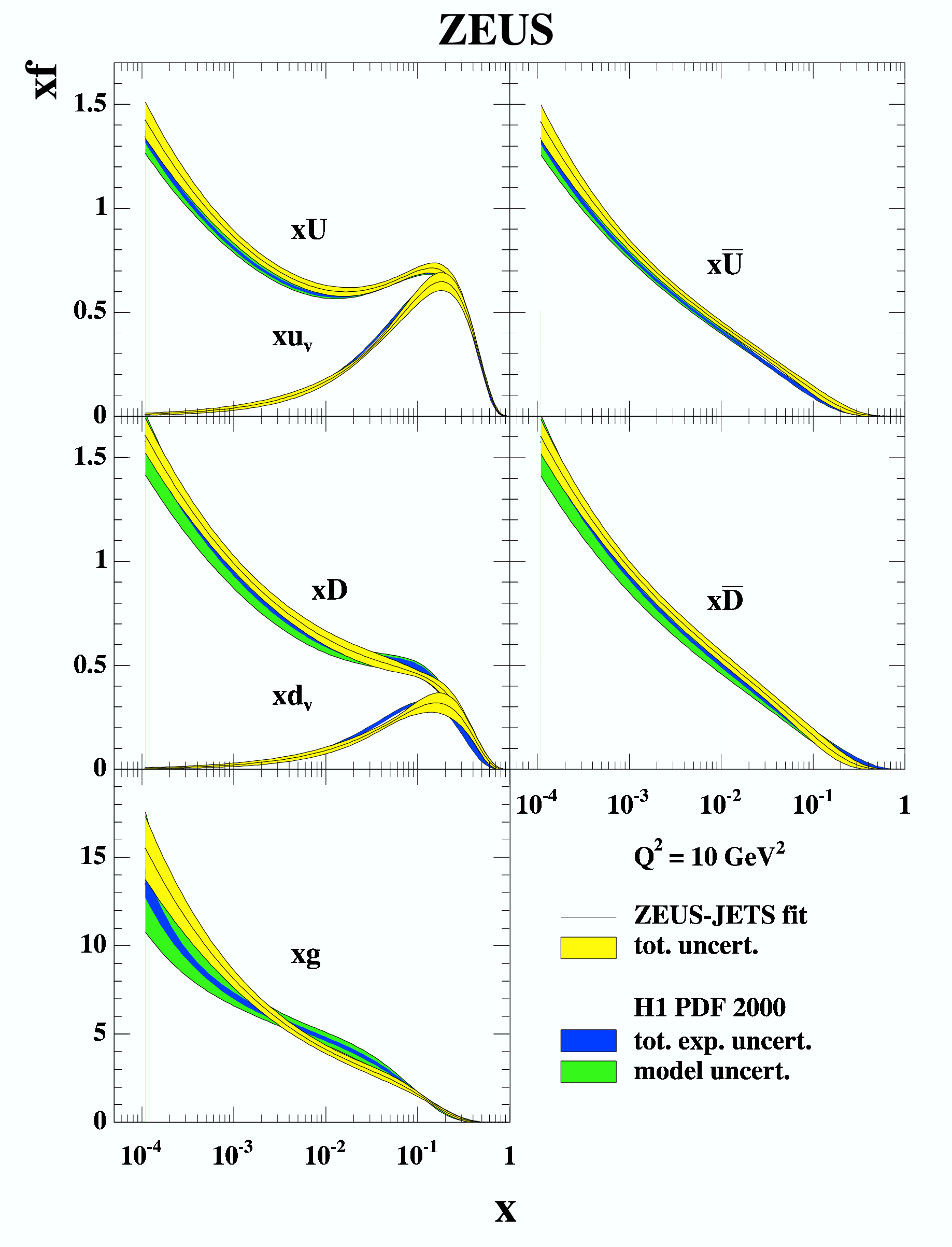}}
   \end{picture}
   \caption{Left: comparison of pdfs from ZEUS-JET and H1 PDF 2000 analyses. 
   Right: 
   experimental uncertainties of the gluon pdf of the ZEUS-JET fit (central
   error bands) compared to the uncertainties obtained without including the 
   jet data (outer error band).
}
   \label{fig:gluonerror}
   \end{center}
   \end{figure}

This led the ZEUS collaboration to perform a pdf analysis
 based on one experiment only by fitting the ZEUS
inclusive DIS NC and CC data together with di-jet data measured in photoproduction
and inclusive jets in the Breit frame measured in
 $ep$ DIS~\cite{Chekanov:2005nn} (ZEUS-JETS fit).
The resulting pdfs, presented in Fig.~\ref{fig:gluonerror}, left, show  
indeed a reduction of the uncertainty of the gluon 
density at medium and high $x$ (Fig.~\ref{fig:gluonerror}, right).

Other  approaches have been used in earlier H1 and ZEUS analyses of
inclusive DIS data
 which differ mainly in the choice of input data used, the handling
 of systematic errors, the parameterisations at $Q_0^2$, and the treatment
 of heavy quarks.

 A special case is the H1 NC analysis~\cite{Adloff:2000qk}, based
 mainly  on H1 $ep$ NC data and some BCDMS $\mu p$ high $x$ data.
Here the primary purpose was
 a determination
 of the gluon density $g(x)$
 and of the strong
coupling constant $\alpha_s$.
 For this reason,
 besides $g(x)$ only two
 functions were parametrised  at  $Q_0^2$,
  one for the
 valence
 and one for the sea quark contribution, with small corrections.

 The  H1 2000 pdf fit~\cite{Adloff:2003uh}, which includes in addition
 the H1 CC and the BCDMS $\mu d$ data,\footnote{The fit was also performed 
 with H1 data alone}
 determines $g(x)$ and also the
 four up and down combinations
  $U = u+c,\;\; \bar U = \bar u + \bar c,\;\; D = d + s,\;$ and
  $\bar D = \bar d + \bar s\;\;$ from which the valence densities
  $u_v = U - \bar U$ and $d_v = D - \bar D$ are derived.
  This fit is compared with the ZEUS-JETS fit in Fig.~\ref{fig:gluonerror},
  left. The results are consistent, but differences are visible in the
  gluon distribution.
 
 The previous ZEUS analysis~\cite{Chekanov:2002pv} (``ZEUS-S'') uses
 ZEUS NC data, $\mu p$ and $\mu d$ data from BCDMS, NMC and E665,
  and CCFR $\nu Fe$ data. Results on $g(x),\; u_v(x)$, $d_v(x),\;$
 the total sea and $\bar d - \bar u\;$ are given.

\begin{figure}[htb] \unitlength 1pt
   \begin{center}
   \begin{picture}(150.,190.)
     \put(-72,-8){\includegraphics[width=0.75\textwidth]{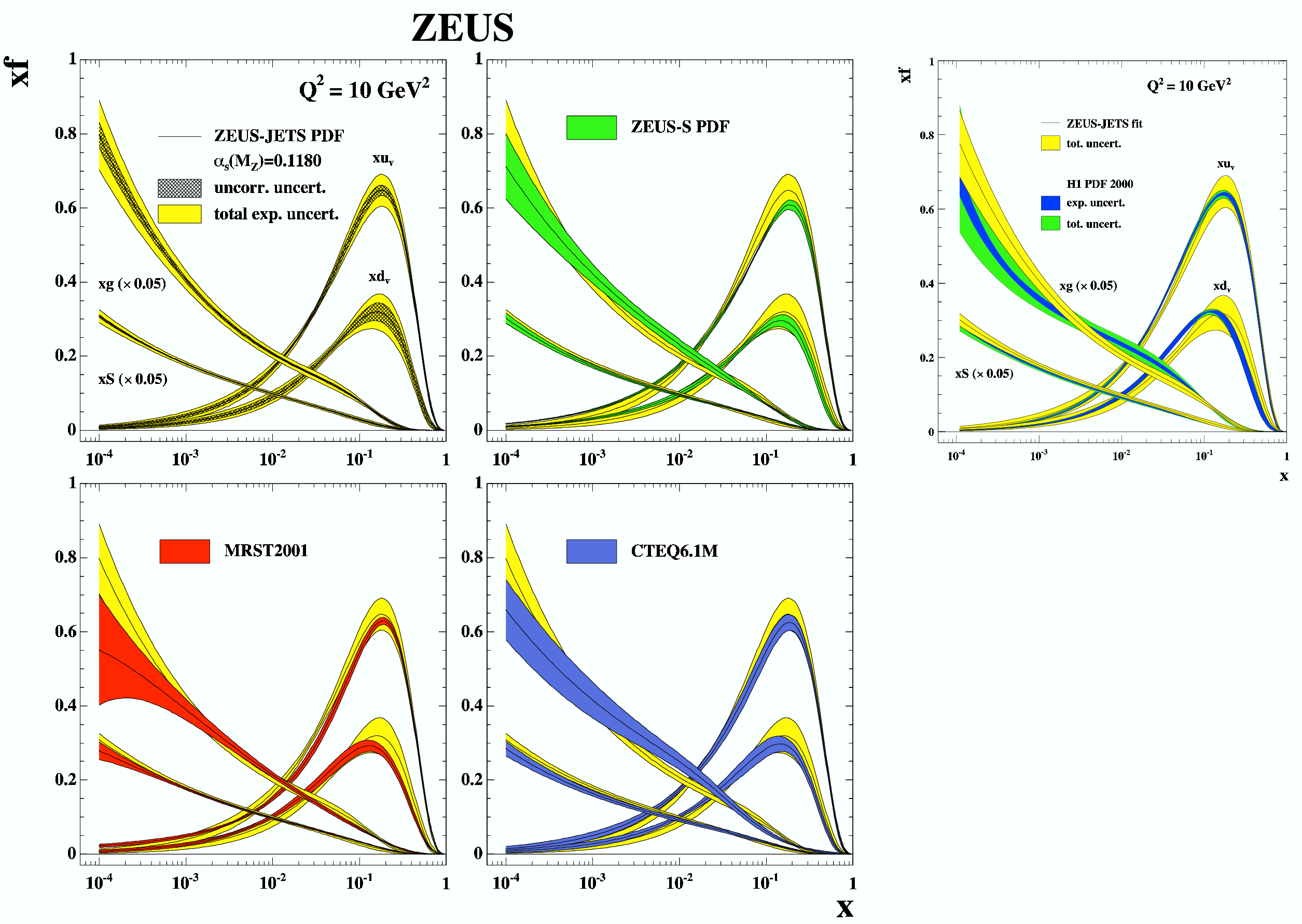}}
   \end{picture}
   \caption{Valence quark, sea and gluon distributions of different
   NLO analyses. Top row from left to right:
   ZEUS-JETS,ZEUS-S and ZEUS-JETS together with H1 PDF 2000.
   Bottom row: ZEUS-JETS with MRST 2001~\cite{Martin:2001es} (left)
    and CTEQ6.1M~\cite{Pumplin:2002vw}(right).
}
   \label{fig:jeth1glob}
   \end{center}
   \end{figure}

 In Fig~\ref{fig:jeth1glob},
 the densities 
  of valence quarks, the sea and the gluon
  are compared for the fits ZEUS-JET,
  ZEUS-S, H1 PDF 2000 and the global analyses
  MRST2001~\cite{Martin:2001es} and CTEQ6.1M~\cite{Pumplin:2002vw}.
 In general there is consistency,
  but one notices that the ZEUS-JET gives lower g(x) at $x \approx 0.01$
  and higher valence $u$ and $d$ densities than the other fits.

\section{Charm and Beauty Contribution to Proton Structure Functions}
\label{sec:beau}

There are different schemes to treat the heavy quark contributions
and thresholds in pQCD pdf fits. The H1 NC analysis~\cite{Adloff:2000qk}
uses the ``massive" scheme, which is particularly useful at low $Q^2$.
In this case the boson gluon fusion contribution 
(Fig.~\ref{fig:graphs}, left) is generated at LO. In the ``massless scheme",
 favoured for high $Q^2$ and
far above the heavy quark production thresholds, the heavy quarks appear
as part of the sea. Interactions equivalent to the boson gluon
graph appear at NLO. This method is used in the 
 H1 2000 pdf fit~\cite{Adloff:2003uh}. The ZEUS-S and ZEUS-JET fits
 use an interpolation between these approaches, a ``variable flavour number
 scheme"~\cite{Thorne:1997ga}.
 
 At HERA energies there are substantial charm and beauty contributions
 to the proton structure functions. The charm contribution to $F_2$ is
 typically 20\% to 30\%, the beauty contribution was measured 
 to be in the range 0.3\% to 3\% (see~\cite{Bell}).
 These numbers suggest to make direct use 
 of the experimental information on charm and beauty production
 in the pQCD pdf analyses.
 
 \section{Conclusion}

  The structure function $F_L$ provides an important constraint
  on $g(x)$ and a consistency check for QCD analyses. No direct
  measurements exist yet at HERA energies.
  
\noindent
 A new approach based on the observation of jets provides new inclusive
 DIS data at high~$x$.
 
\noindent
 The CC data of HERA allow for a separate determination of $u$ and $d$ pdfs,
 free of nuclear corrections.
 
\noindent
 Data on di-jet production can improve pdf-fits at medium $x$.

\noindent
 The new available data on charm and beauty production may provide
 direct input for the heavy flavour treatment in pQCD analyses.

\noindent
HERA II will strongly improve the precision of high $Q^2$ data.

\section*{Acknowledgements}  I am grateful to Emmanuelle Perez and Alex Tapper
for carefully reading the draft.

\end{document}